\documentclass[preprint,12pt]{elsarticle}

\usepackage{epsfig}
\usepackage{graphicx}
\usepackage{amsmath}
\usepackage{rotating}

\newcommand{\oh}{\frac{1}{2}}

\newcommand{\be}{\begin{eqnarray}}
\newcommand{\ee}{\end{eqnarray}}
\newcommand{\ba}{\begin{array}}
\newcommand{\ea}{\end{array}}
\newcommand{\bt}{\begin{tabular}}
\newcommand{\et}{\end{tabular}}
\newcommand{\btab}{\begin{table}}
\newcommand{\etab}{\end{table}}
\newcommand{\bfig}{\begin{figure}}
\newcommand{\efig}{\end{figure}}
\newcommand{\bc}{\begin{center}}
\newcommand{\ec}{\end{center}}
\newcommand{\bi}{\begin{itemize}}
\newcommand{\ei}{\end{itemize}}

\newcommand{\Eq}[1]{Eq.~(\ref{#1})}

\newcommand{\coma}{\textrm{, }}

\newcommand{\tw}{\textwidth}
\newcommand{\ig}[1]{\includegraphics[width=#1\tw]}

\begin{document}

\begin{frontmatter}

\title{A modular synthetic device to calibrate promoters}

\author[catedra,poli]{D. Gamermann}\ead{daniel.gamermann@ucv.es}
\author[poli]{A. Montagud}
\author[tarragona]{P. Aparicio}
\author[malaga]{E. Navarro}
\author[pinar]{J. Triana}
\author[malaga]{F. R. Villatoro}
\author[poli]{J.~F.~Urchuegu\'ia}
\author[poli]{P.~Fern\'andez de C\'ordoba}

\address[catedra]{C\'atedra Energesis de Tecnolog\'ia Interdisciplinar, Universidad Cat\'olica de Valencia San Vicente M\'artir, \\ Guillem de Castro 94, E-46003, Valencia, Spain.}
\address[poli]{Instituto Universitario de Matem\'atica Pura y Aplicada, Universidad Polit\'ecnica de Valencia,\\  Camino de Vera 14, 46022 Valencia, Spain.}
\address[tarragona]{Departament de Qu\'imica F\'isica i Inorg\`anica, \\ Universitat Rovira i Virgili, 43007, Tarragona, Spain.}
\address[malaga]{Departamento de Lenguajes y Ciencias de la Computación,\\E.T.S.I Industriales, Universidad de M\'alaga,\\
  Campus El Ejido, S/n 29013, M\'alaga, Spain.}
\address[pinar]{Departamento de Qu\'imica,\\ Universidad Pinar del R\'io ``Hermanos Sa\'iz Montes de Oca'',\\ Mart\'i 270, 20110, Pinar del R\'io, Cuba.}


\begin{abstract}
In this contribution, a design of a synthetic calibration genetic circuit to characterize the relative strength of different sensing promoters is proposed and its specifications and performance are analyzed via an effective mathematical model. Our calibrator device possesses certain novel and useful features like modularity (and thus the possibility of being used in many different biological contexts), simplicity, being based on a single cell, high sensitivity and fast response. To uncover the critical model parameters and the corresponding parameter domain at which the calibrator performance will be optimal, a sensitivity analysis of the model parameters was carried out over a given range of sensing protein concentrations (acting as input). Our analysis suggests that the half saturation constants for repression, sensing and difference in binding cooperativity (Hill coefficients) for repression are the key to the performance of the proposed device. They furthermore are determinant for the  sensing speed of the device, showing that it is possible to produce detectable differences in the repression protein concentrations and in turn in the corresponding fluorescence in less than two hours. This analysis paves the way for the design, experimental construction and validation of a new family of functional genetic circuits for the purpose of calibrating promoters.
\end{abstract}


\end{frontmatter}


\section{Introduction}

One of the fundamental principles of synthetic biology is the construction of biological standardized parts and devices which are interchangeables. A proper characterization of these parts and devices appears as a key issue in order to make them reusable in a predictive way.  In the recent past scientists have witnessed several initiatives towards the design and fabrication of synthetic biological components and systems as a promising way to explore, understand and obtain beneficial applications from nature. For instance, in the post genomic era one of the most fascinating challenges scientists are facing is to understand how the phenotypic behaviour of  living cells arise out of the properties of their complex network of signalling proteins. While the interacting biomolecules perform many essential functions in these systems, the underlying design principles behind the functioning of such intracellular networks still remain poorly understood \cite{elowitz,becskei}. Several initiatives have been reported in this line of thought to uncover some key  working principles of such genetic regulatory networks via quantitative analysis of some relatively simple, experimentally well characterized, artificial genetic circuits. It has been shown that custom made gene-regulatory circuits with any desired property can be constructed from simple regulatory elements \cite{monod}. These properties include bistability, multistability or oscillatiory behaviour of genetic circuits in various microorganisms such as bacteriophage switch \cite{ptashne} or the cyanobacterium circadian oscillator \cite{ishiura}. As one example, the genetic {\it toggle switch}, a synthetic, bi-stable gene-regulatory network in {\it Escherichia coli}, was shown to provide a simple theory that uncovers the conditions necessary for bi-stability \cite{gardner,stricker}. Further, artificial positive feedback loops (PFLs) have been used as genetic amplifiers in order to enhance the responses of weak promoters and in the creation of eukaryotic gene switches \cite{becskei2}. Sayut et al. demonstrated the construction and directed evolution of two PFLs based on the LuxR transcriptional activator and its cognate promoter, Pluxl \cite{sayut}. These circuits may have application in metabolic engineering or gene therapy that requires inducible gene expressions \cite{weber,walz}.

The desired performance of these synthetic networks and in turn the resultant phenotype is strongly dependent on the expression level of the corresponding genes, which is further controlled by several factors such as promoter strength, cis- and trans-acting factors, cell growth stage, the expression level of various RNA polymerase-associated factors and other gene-level regulation characteristics \cite{gardner,becskei}. Thus, one important ingredient to elucidate gene function and genetic control on phenotype would be to have access to well-characterized promoter libraries. These promoter libraries would be in turn useful for the design and construction of novel biological systems. There have been several initiatives to control gene expression through the creation of promoter libraries \cite{kumar,santos}. Alper et al., \cite{alper} have reported a methodology to develop a completely characterized, homogeneous, broad-range, functional promoter library with the demonstration of its applicability to analysis of genetic control.

Since Miller published \cite{miller} a proposal for a measurement standard for $\beta$-galactosidase assays, yet much work has been done with no conclusive standard being established \cite{liang,smolke,khlebnikov}. The main goal in calibration is measuring a {\it query} value up to an established {\it standard}. A good {\it device} should be unique, reliable and easy to use; additionally it should circumvent, to all possible extent, any noise that could alter the measurement. Recently a methodology \cite{kelly} has been reported to characterize the activity of promoters by using two different cell strains. In the present study we propose the use of a synthetic gene regulatory network as a framework to characterize different promoter specifications by using a single-cell strategy. In this context characterization stands for evaluating the parameters of a {\it query} promoter as compared to a standard promoter acting as a “scale”.  The proposed device, the promoter “calibrator”, works on the principle of comparing a specific input signal which will be sensed by promoters of different sensing strengths and, as an output, produces fluorescence of specific colours which allows quantifying the relative strength of the promoters. Analyses were carried out in order to find out relevant model parameters and the corresponding range of model parameter values which are compatible with the performance of this calibrating biological design over a spectrum of given input . 

This contribution is organized as follows: in the first part, ``Design'', the structure and working principle are explained and the mathematical model resulting from the construction is established. In section \ref{sec:numana}, ``Numerical Analysis of the System'', we analyze the dynamics of the model equations in regard to its stability, functional parameter regions and sensitivity or robustness vs. the change in certain key parameter values. In the following section, a proof of concept design is proposed in order to choose the right parameters to actually perform the experimental validation of our concepts and have a system that gives a clear and stable signal that can be interpreted. Finally, the conclusions resulting from our paper are exposed.


\section{Design}

\subsection{Biological principles}

\begin{figure}
\begin{center}
\includegraphics[width=7cm]{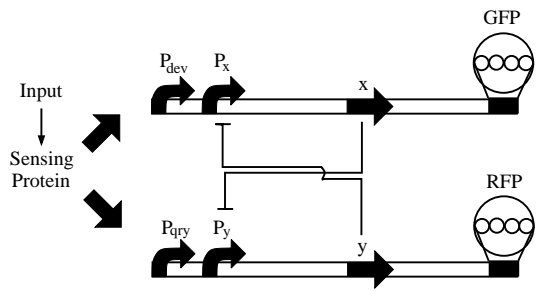}
\caption{Design of the proposed promoter calibrator. It is composed of two promoters (with two parts each: a sensing and a repressed domain) one of the sensing promoters is the {\it device} promoter and the other is the {\it query} promoter. The repressed domains are controled by the two repressors proteins ($x$ and $y$). Each promoter is inhibited by the repressor which is transcribed from the opposing promoter. Fluorescence proteins levels will be proportional to repressor protein levels, which, in turn, will be promoted by the sensing promoters.} \label{fig:diagr}
\end{center}
\end{figure}

Our promoter calibrator is composed of two promoters (each with two parts: a sensing and a repressed domain), two repressors proteins and two fluorescent protein outputs (see Fig. \ref{fig:diagr}). Each promoter is inhibited by the repressor, transcription of which is promoted by the opposing promoter. Fluorescence protein levels will be directly related to repressor protein levels, activated in turn by their sensing promoters. Hence, different sensing strengths will cause a difference in the expression of the fluorescence proteins, detectable by means of single cell fluorescence as changes in the color patterns of the individual cell or cell sample. 

In our scheme, one of the sensing promoters acts as the {\it device} promoter to which the strength of a given {\it query} promoter is quantitatively compared. The main use of this device is to characterize different promoter specifications (sensing affinities and cooperativities) compared to some standard. One of the main usefulness of this design lies in the potential modularity of the system: by changing the sensing part of the promoters, other sensing promoters could be calibrated; this change can be carried out by a simple, straight-forward cloning step. Modularity also boasts the potential of this device as it can be implemented in a potentially unlimited set of systems.	

\subsection{Mathematical model}

The behaviour of the proposed promoter calibrator can be understood via an effective mathematical model. The model is considered to be effective as transcription and translation have been modeled as a lumped reaction. The separation of transcription and translation otherwise involves a response delay.  We seek to classify dynamic behaviors depending upon the change in model parameters and determine which experimental parameters should be fine-tuned in order to obtain a satisfactory performance of our device. 

The time dependent changes in repressor and sensing protein (input) concentrations is shown in equations (\ref{eq1}-\ref{eq3}). Subsequent to the biological design, reporter protein concentrations are directly related to repressor protein concentrations.

\be
\frac{dx}{dt}&=&\alpha_1\frac{\left(\frac{p_s}{k_1}\right)^{n_1}}{1+\left(\frac{p_s}{k_1}\right)^{n_1}}
  \frac{1}{1+\left(\frac{y}{k_y}\right)^{n_y}}-\beta_x x +\gamma_x \label{eq1}, \\
\frac{dy}{dt}&=&\alpha_2\frac{\left(\frac{p_s}{k_2}\right)^{n_2}}{1+\left(\frac{p_s}{k_2}\right)^{n_2}}
  \frac{1}{1+\left(\frac{x}{k_x}\right)^{n_x}}-\beta_y y +\gamma_y \label{eq2}, \\
\frac{dp_s}{dt}&=&-\beta_{p_s} p_s\label{eq3}.
\ee

The {\it device} and {\it query} promoters activate the production of repressor protein $x$ and $y$, respectively, and their concentration is related directly to the concentration of fluorescence proteins. Thus these variables will be treated as equivalent from the modelling point of view. Parameters $\alpha_1$ and $\alpha_2$ represent the effective rate of synthesis of repressor proteins $x$ and $y$, respectively; $\alpha$ is a lumped parameter that takes into account the net effect of various activities such as RNA polymerase binding, RNA elongation and termination of transcript, ribosome binding and polypeptide elongation and will be modified by repression and sensing effects. The $\beta_x$, $\beta_y$ and $\beta_{p_s}$ are the degradation constants of repressor protein $x$, repressor protein $y$ and sensing protein $p_s$, respectively. The sensing protein concentration $p_s$ will depend on the sensed input, will be easy to change in a given experiment and is used as the main input variable in our calibrator experiments. It is important to note that a slow rate of degradation is assumed for the sensing protein, implying a nearly constant level over a reasonable experimental time interval. Basal level rates of synthesis of proteins $x$ and $y$ are denoted by $\gamma_x$ and $\gamma_y$, respectively.

Repressor and sensing responses are assumed to follow Hill equation dynamics: promoter-binding monomers form multimers by positive allosterism and attach to its cognate promoter with saturating behaviour. Binding cooperativities are described by Hill coefficients $n_x$ and $n_y$ for repressor domains corresponding to $x$ and $y$ respectively, and $n_1$ and $n_2$ for sensing domains corresponding to {\it device} and {\it query} promoter respectively. The extent of the saturation rate is described by half saturation constants or Michaelis constants, denoted by parameter $k_x$ and $k_y$ for repressor domains corresponding to $x$ and $y$ respectively and $k_1$ and $k_2$ for sensing domains corresponding to {\it device} and {\it query} promoter respectively. The total number of promoter sites is assumed to be conserved and the total concentration of both promoters is chosen to be identical.

In our construction, the crossrepressing part will be kept unchanged while different sensing domains may be attached to it. The aim is to establish a protocol to accurately quantify differences between the sensing promoter parameters ($\alpha_{1,2}$, $k_{1,2}$). Crossrepression parameters ($k_{x,y}$, $\beta_{x,y}$ and $n_{x,y}$) are structural parameters that must be chosen in such a way that the fluorescence response of the system gives us stable, sensitive and robust indication about the quantitative relations between the sensing promoter parameters. The dynamic analysis of the system will help us to take the right decisions on which are the most appropriate values for these structural parameters. The next sections are devoted to the dynamical analysis in order to determine the sensitivity and robustness of the system for different ranges of the structural parameters.

The commercial software package Mathematica (Wolfram), was used for model development and simulation. In the numerical calculations we have used the following dimensionless variables:

\be
X&=&\frac{x}{k_x} \\
Y&=&\frac{y}{k_y} \\
\tau&=&t\beta_x \\
\bar{\alpha}_{1,2}&=&\frac{\alpha_{1,2}}{\beta_x k_{x,y}}\\
\bar{\gamma}_{x,y}&=&\frac{\gamma_{x,y}}{\beta_x k_{x,y}}
\ee
therefore, the units in the plots of the figures in this work are given in units of $k_x$ or $k_y$ for the $x$ and $y$ repressor proteins concentrations and time in units of $\frac{1}{\beta_x}$. For the adimensional variables, Equations (\ref{eq1}-\ref{eq2}) take the form:

\be
\frac{dX}{d\tau}&=&\bar{\alpha}_{1}\frac{\left(\frac{p_s}{k_1}\right)^{n_1}}{1+\left(\frac{p_s}{k_1}\right)^{n_1}}
  \frac{1}{1+Y^{n_y}}- X +\bar{\gamma}_{x} \label{eq1ad}, \\
\frac{dY}{d\tau}&=&\bar{\alpha}_{2}\frac{\left(\frac{p_s}{k_2}\right)^{n_2}}{1+\left(\frac{p_s}{k_2}\right)^{n_2}}
  \frac{1}{1+X^{n_x}}-R Y +\bar{\gamma}_{y} \label{eq2ad},
\ee
where $R$ is the ratio $\frac{\beta_y}{\beta_x}$.


\section{Numerical analysis of the system}\label{sec:numana}

The simplifying assumption of considering sensing proteins for which the degradation constant $\beta_{p_s}$ is much smaller than the rest ($\beta_{p_s}\ll \beta_x,\beta_y$) was made in order to classify the possible dynamic scenarios of our model. Given this assumption, in a first order of approximation we have,

\be
\frac{dp_s}{dt}&=&-\beta_{p_s}p_s\approx 0.
\ee

In such approach, the concentration of sensing protein $p_s$ is constant during the evolution time of the rest of the internal variables of the system. This assumption leads to a system of two autonomous coupled non-linear ordinary differential equations dependent on the variables $x$ and $y$, eqs. (\ref{eq1ad}-\ref{eq2ad}), in which $p_s$ is fixed although it can be easily changed within a given experiment. This is not true for the rest of parameters which are more difficult to modify in a given experiment.
This approximation transforms the system into:

\be
\frac{dX}{d\tau}&=&\bar{\alpha}^\prime_1\frac{1}{1+Y^{n_y}}-X +\bar{\gamma}_x \label{eq5}, \\
\frac{dY}{d\tau}&=&\bar{\alpha}^\prime_2\frac{1}{1+X^{n_x}}-R Y +\bar{\gamma}_y \label{eq6}.
\ee
where the new parameters $\bar{\alpha}^\prime_i$ (effective transcription factors) are given by the following expression:

\be
\bar{\alpha}^\prime&=&\alpha \frac{\left(\frac{p_s}{k}\right)^{n}}{1+\left(\frac{p_s}{k}\right)^{n}} \label{eq7}.
\ee

\begin{figure}
\begin{center}
\includegraphics[width=7cm]{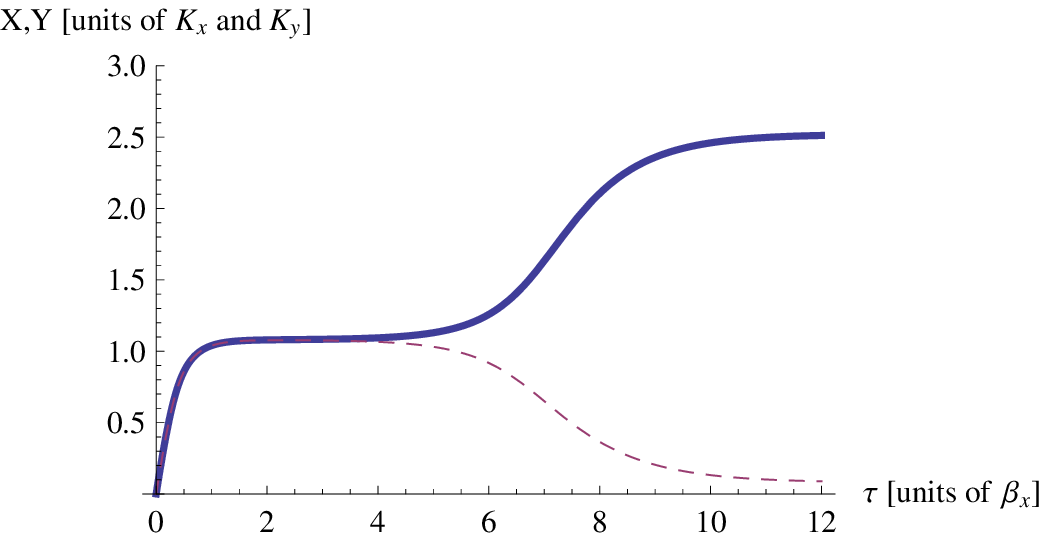} \\
\includegraphics[width=7cm]{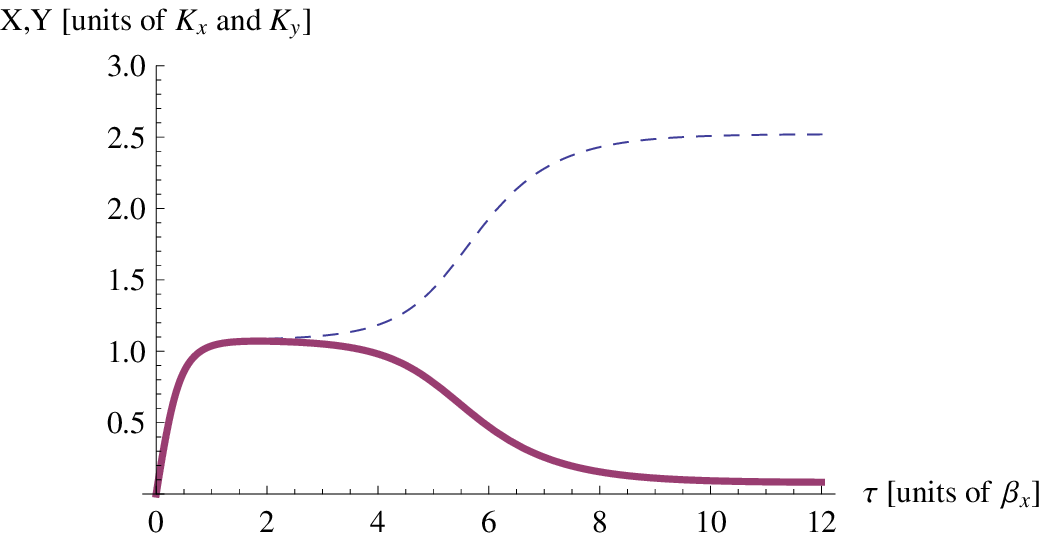}
\caption{Typical response of the proposed promoter calibrator. In the upper figure the concentration of the $x$ protein (solid line) in the steady state is higher while in the figure below the concentration of the $y$ protein (dashed line) is higher.} \label{fig:concentr}
\end{center}
\end{figure}

In the limit in which the constants $k_{x,y}$, $\beta_{x,y}$, $\gamma_{x,y}$ are equal, this equations describe the biological equivalent of an electronic {\it comparator}, that is, a device which compares two voltages or currents and switches its output to the larger signal. In the biological equivalent, our comparator would select for the larger of the two $\bar{\alpha}$'s, as exemplified in Fig. \ref{fig:concentr}, which represent the evolution of the system for the cases in which the {\it query} promoter has a higher and lower effective transcription factor compared to the device promoter, respectively. 	

In any case, our aim is to construct a device, termed a {\it calibrator}, which not only selects the stronger affinity but also allows quantifying the relative strength of both promoters. Although the comparator is a fundamental part of this device, a deeper understanding of the dynamics of the system is required for its application as a calibrator device in real biological environments.	

\subsection{Dynamic analysis of the calibrator}

The dynamical analysis of the system given by Eqs. (\ref{eq5}-\ref{eq6}) requires the determination of its steady state solutions and their linear stability. The steady states $(x_{ss},y_{ss})$ are given by the intersection of the null clines:

\be
F_1(X,Y)&=&\bar{\alpha}^\prime_1\frac{1}{1+Y^{n_y}}-X +\bar{\gamma}_x=0 \coma \label{eq8}
\ee
and

\be
F_2(X,Y)&=&\bar{\alpha}^\prime_2\frac{1}{1+X^{n_x}}-R Y +\bar{\gamma}_y=0 . \label{eq9}
\ee

The analytical solution of Eqs. (\ref{eq8}-\ref{eq9}) cannot be obtained, hence numerical methods must be used. The linear stability of the steady states is determined by the sign of the eigenvalues of the Jacobian matrix, 

\be
M&=&\left(\ba{cc} \frac{\partial F_1}{\partial X} & \frac{\partial F_1}{\partial Y} \\
                  \frac{\partial F_2}{\partial X} & \frac{\partial F_2}{\partial Y} \ea\right)_{X=X_{ss},Y=Y_{ss}}\label{eq10}
\ee

which are given by

\be
\lambda_\pm&=&-\frac{1+R}{2}\pm\oh\sqrt{(R-1)^2+4\Delta} \coma \label{eq11}\\
\Delta&=&\frac{n_x n_y(X_{ss}-\bar{\gamma}_x)(\bar{\alpha}^\prime_1+\bar{\gamma}_x-X_{ss})(Y_{ss}R-\bar{\gamma}_y)(\bar{\alpha}^\prime_2+\bar{\gamma}_y-Y_{ss}R)}{\bar{\alpha}^\prime_1\bar{\alpha}^\prime_2X_{ss}Y_{ss}}. \label{eq12}
\ee

From the analysis of the previous equations (\ref{eq8}-\ref{eq9}), we deduce that, for the positive steady state solutions ($X_{ss}>0$ and $Y_{ss}>0$), the following mathematical constraints hold: $\bar{\alpha}^\prime_1>X_{ss}-\bar{\gamma}_x>0$ and $\bar{\alpha}^\prime_2>Y_{ss}R-\bar{\gamma}_y>0$, respectively. Thus, taking into account (\ref{eq11}-\ref{eq12}), we observe that $\Delta>0$ and $\lambda_-$ is always negative. However, $\lambda_+$ can be either negative, for $\Delta>R$, or positive, for $\Delta<R$, resulting in either stable nodes (sinks) or unstable saddles, respectively. The condition $\Delta=R$ is satisfied at certain critical values of the parameters at which precisely one of the steady state solutions of the system changes its stability.  	

In order to highlight the specific aspects of the calibrator dynamics, we will in the following sections consider a number of special cases. Specifically we will examine the (fully) symmetrical calibrator, $\bar{\alpha}^\prime_1=\bar{\alpha}^\prime_2=\bar{\alpha}^\prime$, $n_x = n_y = n$, $k_x = k_y = k$, $\beta_x=\beta_y\Rightarrow R=1$ and $\bar{\gamma}_x=\bar{\gamma}_y=\bar{\gamma}$, and the partially symmetrical calibrator, with the same specifications except that $\bar{\alpha}^\prime_1$ and $\bar{\alpha}^\prime_2$ may differ. At the end of the section some general considerations about dynamics of the system in the most general case will made.

\subsection{The fully symmetrical calibrator ($\bar{\alpha}_1^\prime=\bar{\alpha}_2^\prime=\bar{\alpha}^\prime$)}

From the analysis of  Eqs. (\ref{eq8}-\ref{eq9}) it is shown that there is always a fixed point with $Y_{ss}=X_{ss}$ and that there exists a minimum value of $X_m$ such that for parameters resulting in $X_{ss}>X_m$, three steady states exist, otherwise only one. 

\begin{figure}
\begin{center}
\includegraphics[width=7cm]{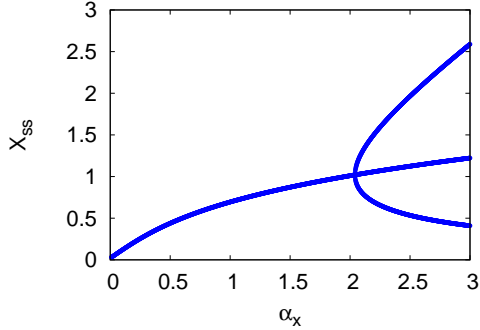} 
\caption{Bifurcation diagram for $X_{ss}$.} \label{fig:bifurc}
\end{center}
\end{figure}

Using $\bar{\alpha}^\prime$ as free parameter and taking fixed values for the rest, i.e., $n$, $R$ and $\bar{\gamma}$, the condition  $\Delta=R=1$, together with \Eq{eq8}, allows to obtain the critical values $\bar{\alpha}^\prime_m$ and $X_m$ that characterize the appearance of the bifurcation, namely:

\be
1&=&\frac{n^2(\bar{\gamma}-X_m)^2(\bar{\gamma}-X_m+\bar{\alpha}^\prime_m)^2}{X_m^2\bar{\alpha}^{\prime 2}_m}
\ee
whose values can be obtained by numerical methods. For example, for $n=2$, $k=80$, $\beta=0.069$ and $\bar{\gamma}=0.1$, yields $\bar{\alpha}^\prime_m=11.24$ and $x_m=81.46$ or, in the dimensionless variables: $X=1.018$ and $\bar{\alpha}^\prime=2.036$. Figure \ref{fig:bifurc}, shows the bifurcation diagram for $X_{ss}$ as function of $\bar{\alpha}$ showing that for $\bar{\alpha}>\bar{\alpha}_m$ there are three steady states.

This analysis shows that the (fully) symmetrical calibrator possesses three fixed points for $\bar{\alpha}^\prime_1>\bar{\alpha}^\prime_m$: a saddle ($\vec{x_M}$) with $X_{ss}=Y_{ss}$, and two sinks, one with $X_{ss}>Y_{ss}$ and another one with $X_{ss}<Y_{ss}$, referred to as $\vec{x}_R$ and $\vec{x}_L$, respectively. This behaviour is typical of the occurrence of a (supercritical) pitchfork bifurcation and bistable behaviour.

\begin{figure}
\begin{center}
\includegraphics[width=7cm]{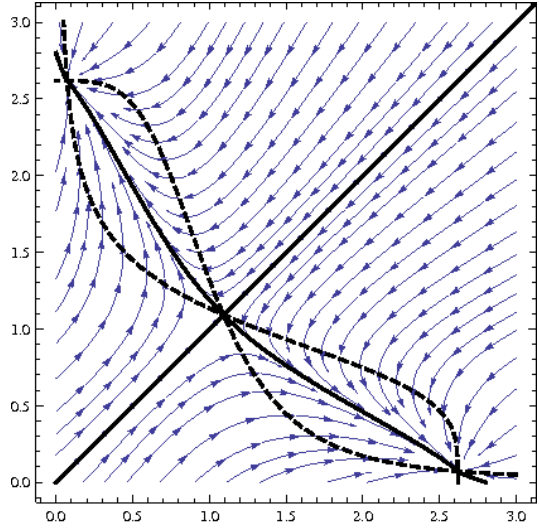} 
\caption{Phase plane, showing the unstable equilibrium point (the point where the two dashed lines touch in the center) and the two steady state solutions (points where the dashed lines touch close to each axis). The arrows show the path the system would do starting from any point in the phase space.} \label{fig:vectorfi}
\end{center}
\end{figure}

Regarding the possible trajectories of the dynamic variables, Figure \ref{fig:vectorfi} illustrates the phase plane of Eqs. (\ref{eq5}-\ref{eq6}), where the steady states are located at the intersection of the null clines eqs.(\ref{eq8}-\ref{eq9}) represented by dashed lines. The solid lines are the stable ($W^S$) and unstable ($W^U$) manifolds of the saddle fixed point $\vec{x_M}$. The stable manifold $W^S$ divides the phase plane in two regions, the first and second octants corresponding to the attraction basins of the sinks $\vec{x}_R$ and $\vec{x}_L$, respectively. Different possible trajectories in the phase plane are depicted for a given number of initial conditions, where the arrows indicate the flow direction.

In a calibrator experiment the initial value of the repressor protein concentrations $x$ and $y$ would be zero and hence the phase plane trajectories would depart from the origin in Figure 4. For values of $\bar{\alpha}^\prime$ larger than $\bar{\alpha}^\prime_m$, the system becomes unpredictable, as small perturbations in the trajectories would potentially push the system into any of the attraction basins of the sinks $\vec{x_R}$ and $\vec{x_L}$.

\subsection{The partially symmetrical calibrator}

We consider now the more general scenario in which $\bar{\alpha}^\prime_1$ and $\bar{\alpha}^\prime_2$ may differ being the rest of variables equal ($n_x = n_y = n$, $k_x = k_y = k$, $\beta_x=\beta_y=\beta$ and $\bar{\gamma}_x=\bar{\gamma}_y=\bar{\gamma}$). The condition $\Delta=R$ which characterizes the occurrence of the pitchfork bifurcations now reads:

\be
1&=&\frac{n^2(\bar{\gamma}-X_{ss})(\bar{\gamma}-Y_{ss})(\bar{\gamma}-X_{ss}+\bar{\alpha}^\prime_1)(\bar{\gamma}-Y_{ss}+\bar{\alpha}^\prime_2)}{X_{ss}Y_{ss}\bar{\alpha}^\prime_x\bar{\alpha}^\prime_y}\label{eq:bifurc}
\ee
that shall be solved together with Eqs. (\ref{eq8}-\ref{eq9}) for the fixed points of the system. 

Fig. \ref{fig:diffal} shows the result of the numerical simulation of the resulting system of equations (with initial conditions $X=Y=0$) by slightly changing the value of $\bar{\alpha}_2^\prime$ with respect to $\bar{\alpha}_1^\prime$. The figure shows the results of different simulations for $\bar{\alpha}_1^\prime=$3.0, $n=3$ and $\bar{\alpha}_2^\prime=\epsilon\bar{\alpha}_1^\prime$ with $\epsilon=$0.5, 0.6, 0.7, ..., 1.0, ..., 1.5. The results for $\epsilon<1$ are the points in the right down corner of the plot. One can see that these points positions are very insensitive to the value of $\bar{\alpha}_2^\prime$. There is only one point in the center of the plot, which corresponds to $\bar{\alpha}_1^\prime=\bar{\alpha}_2^\prime$, it is the unstable saddle, and small perturbations in the system will drive the system away from this solution to either of the other two steady state solutions. Once $\epsilon>1$, the system goes to the solutions where $Y_{ss}>X_{ss}$ which are represented by the points in the upper left corner. For these points the maximum value of $\bar{\alpha}^\prime$ is growing and one can observe that the solution is sensitive to this value. So the steady state solution into which the system falls is only sensitive to the bigger value between $\bar{\alpha}_1^\prime$ and $\bar{\alpha}_2^\prime$ and changes in the smaller among these two parameters has no sensible effect in the final solution.

\bfig
\bc
\ig{0.65}{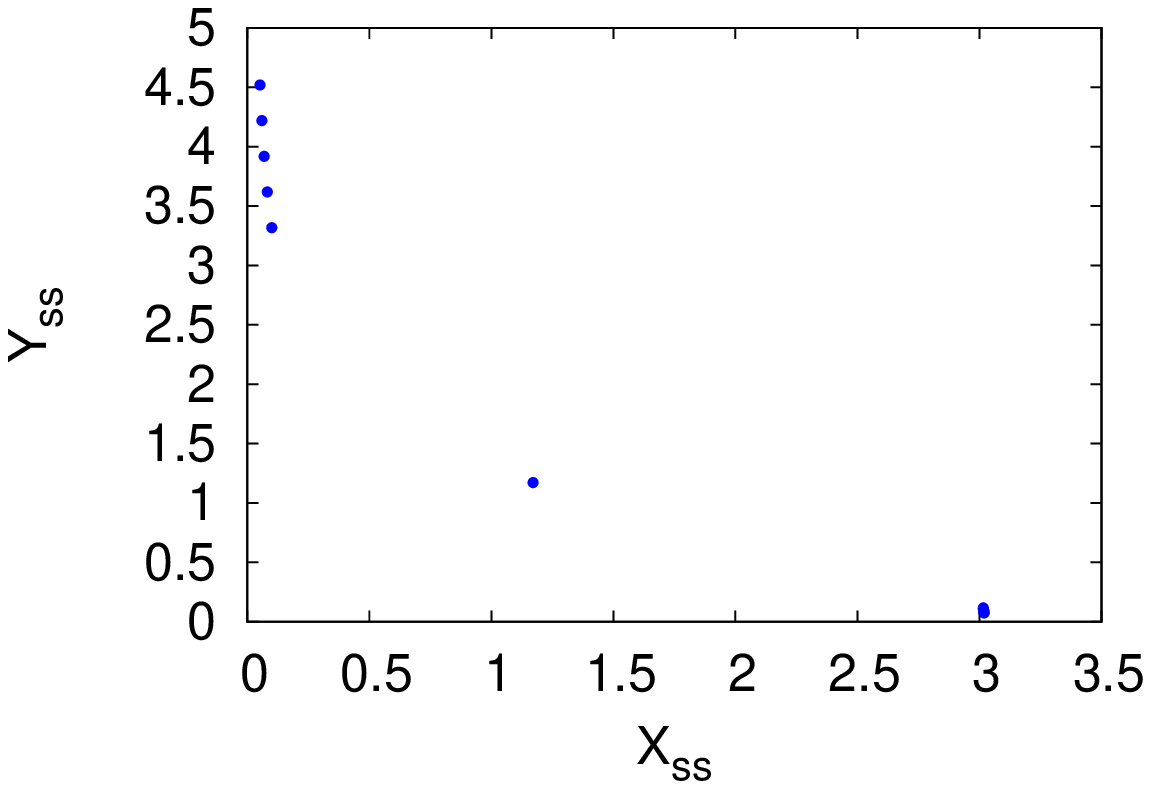}
\caption{Results for the simulation of the partially symmetric calibrator. On the right, close to the x-axis ($\bar{\alpha}^\prime_2<\bar{\alpha}^\prime_1$ for these points), there are many points at the same position, showing that, for the parameter region where a bifurcation happens, the solution is insensitive to the value of the weakest between the two $\bar{\alpha}^\prime$s.}\label{fig:diffal}
\ec
\efig

For the case in which the calibrator falls within the region of bistability, if $\bar{\alpha}^\prime_2<\bar{\alpha}^\prime_1$ the orbits departing from the origin of Fig. \ref{fig:diffal} would fall within the attraction basin of solution $\vec{x_R}$. It is nevertheless observed that $\vec{x_R}$ is quite insensitive to the actual $\bar{\alpha}^\prime_2 / \bar{\alpha}^\prime_1$ ratio. In consequence, the system would show a stable but rather insensitive response to different {\it query} promoters. On the other hand, if $\bar{\alpha}^\prime_1<\bar{\alpha}^\prime_2$, the orbits departing from the origin would fall within the attraction basin of solution $\vec{x_L}$, which changes appreciably as a function of the $\bar{\alpha}^\prime_2 / \bar{\alpha}^\prime_1$ ratio. Thus the system would not only be stable, but also rather sensitive to changes in the effective {\it query} promoter affinity. It should be kept in mind that the sensing protein concentration, $p_s$, can be used to modify $\bar{\alpha}^\prime_1$, $\bar{\alpha}^\prime_2$, which changes from unity to $\bar{\alpha}_{1,2}$ as $p_s$ changes from zero to infinity and therefore the ratio $\bar{\alpha}^\prime_2 / \bar{\alpha}^\prime_1$ changes with $p_s$.

We can also define the fluorescence ratio as the ratio of $X/Y$ if $X<Y$ and $Y/X$ if $Y>X$. This will be the intensity ratio of the two fluorescences once the system reaches stability. In Fig. \ref{fig:ratio1} we show a plot of this ratio for different values of $\bar{\alpha}^\prime_2 / \bar{\alpha}^\prime_1$. This ratio grows until it reaches its maximum when $\bar{\alpha}^\prime_2=\bar{\alpha}^\prime_1$ and then it decreases. Another observation about this parameter is that the bigger $\bar{\alpha}^\prime_1$ is, the less sensible to the ratio $\bar{\alpha}^\prime_2/\bar{\alpha}^\prime_1$ the fluorescence ratio will be.

\bfig
\bc
\bt{c}
\ig{0.65}{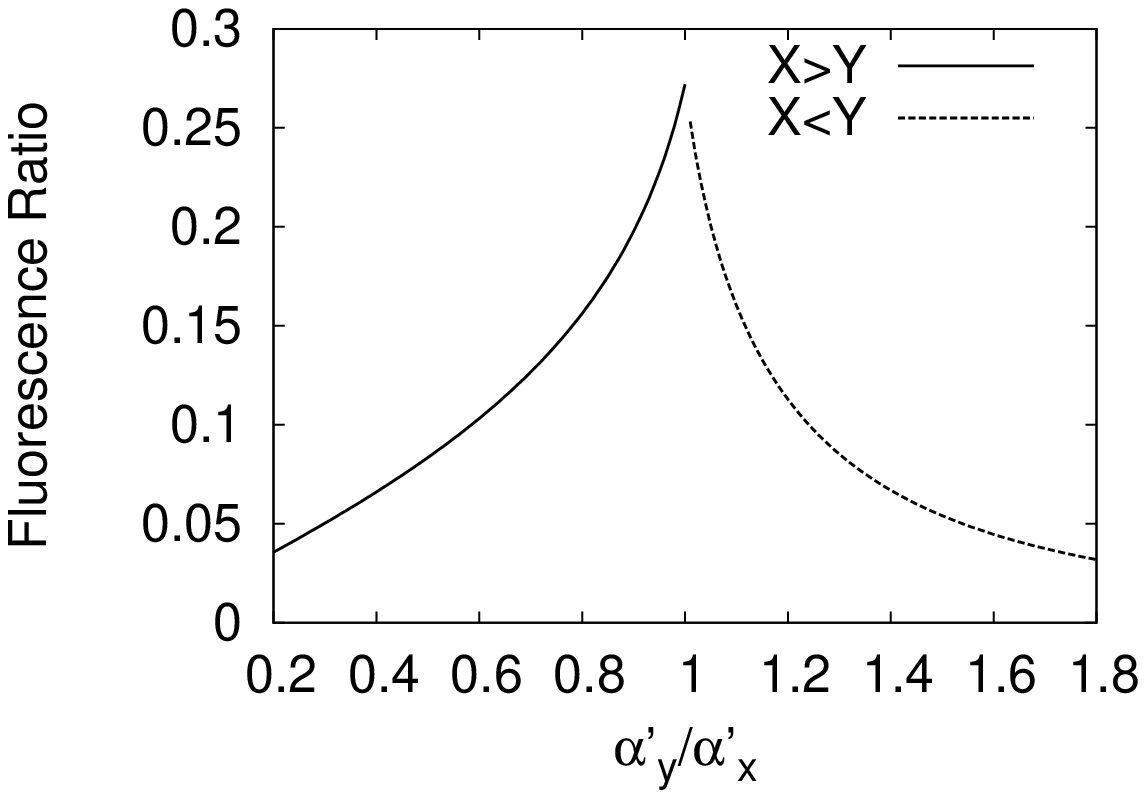}\\
\ig{0.65}{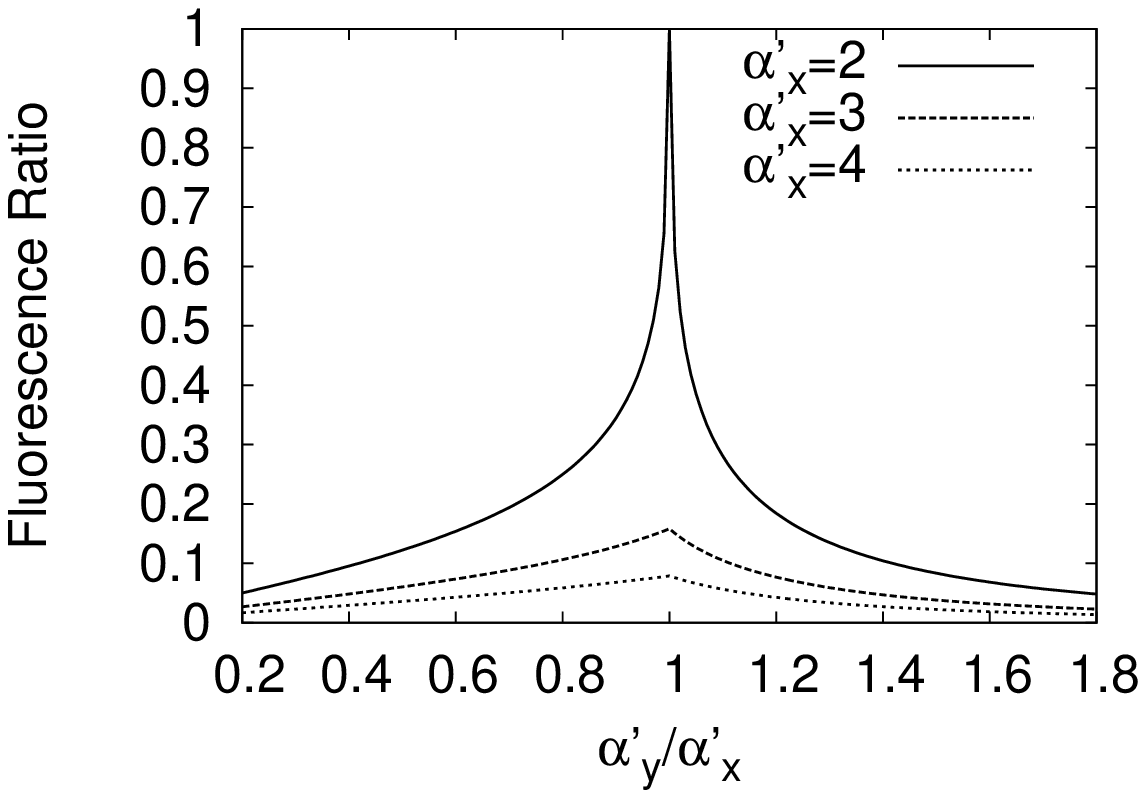}
\et
\caption{Upper plot: Fluorescence ratio for different values of $\bar{\alpha}^\prime_y/\bar{\alpha}^\prime_x$ ($\bar{\alpha}_x^\prime=$2.5). The blue points are solutions where $X>Y$ and in the red points $Y>X$. Lower plot: Fluorescence ratio for different values of $\bar{\alpha}^\prime_y/\bar{\alpha}^\prime_x$ and for different values of $\bar{\alpha}_x^\prime$ (Solid line:$\bar{\alpha}_x^\prime=$2, dashed line:$\bar{\alpha}_x^\prime=$3, dotted line:$\bar{\alpha}_x^\prime=$4).}\label{fig:ratio1}
\ec
\efig

\subsection{The calibrator dynamics in the general case}

The theorem of Andronov and Pontryagin \cite{guckenheimer} states that Eqs. (\ref{eq5}-\ref{eq6}) in the symmetrical case are structurally stable, since every fixed point is hyperbolic (its eigenvalues have a non-null real part) and there are no orbits connecting two saddles (since there is only one). Structural stability implies that the phase plane topology is preserved under small perturbations of the parameters. Hence, the phase plane of Eqs. (\ref{eq5}-\ref{eq6}) in the case that $\bar{\alpha}^\prime_x\approx\bar{\alpha}^\prime_y$, $n_x\approx n_y$, $k_x\approx k_y$, $\beta_x\approx\beta_y$ and $\bar{\gamma}_x\approx\bar{\gamma}_y$, is topologically equivalent to that shown in Fig. 4, meaning that there is a continuous function (homeomorphism) between both phase planes.

Changing the ratio of other structural parameters of the calibrator has similar results as in the partially symmetrical case. For a given range close to the value 1 for the ratio of each parameter ratio ($n_{x/y}$, $\beta_{x/y}$, ...) the bifurcation appears while far from the value 1 the bifurcation cannot be seen. The range is usually bigger, the bigger the values for $\bar{\alpha}^\prime_{1,2}$ are. In Fig. \ref{fig:chabet} we show, as an example, the range where the bifurcation appears for different values of $\beta_x/\beta_y$.

\bfig
\bc
\ig{0.65}{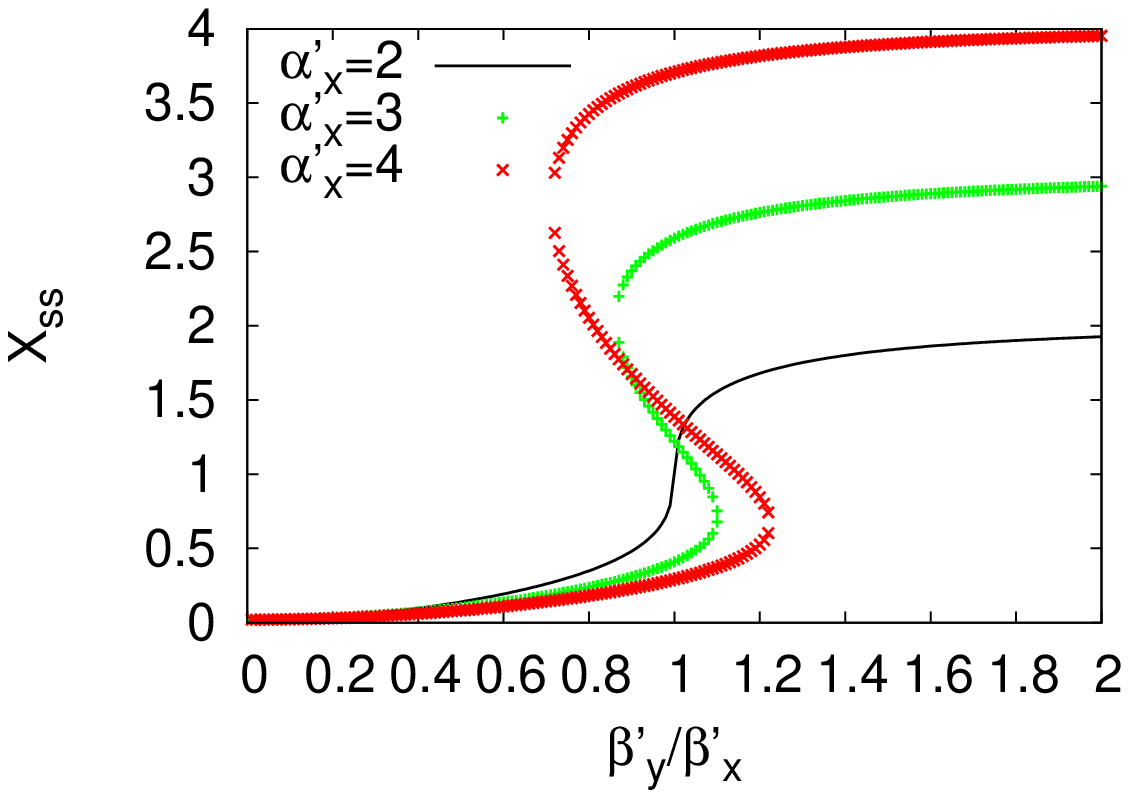}
\caption{(Color online) Position of $X_{ss}$ for different values of $\beta_y/\beta_x$. In the black curve $\bar{\alpha}_x^\prime=\bar{\alpha}_y^\prime=$2, in the blue curve $\bar{\alpha}_x^\prime=\bar{\alpha}_y^\prime=$3 and in the red one $\bar{\alpha}_x^\prime=\bar{\alpha}_y^\prime=$4.}\label{fig:chabet}
\ec
\efig
 
If $R<1$ the orbits departing from the origin ($X=Y=0$) would fall within the attraction basin of solution $\vec{x}_L$, on the other hand if $R>1$ the orbits departing from the origin would fall within the attraction basin of solution $\vec{x}_R$.


\subsection{Calibrator performance analysis: robustness and response time}

In order to use this system to measure the relative strength between two promoters, one should keep in mind two factors. The first important factor is the right choice for the parameters of the repressor proteins and {\it device} promoter in order to have a robust system, that gives a stable response that can be easily interpreted. Second, is the time response of the device, that means, how long does the system needs to reach its steady state solution. 

When the equations are written in the dimensionless form, the parameters $k_x$ and $k_y$ do not appear explicitly, see eqs. (\ref{eq1ad}-\ref{eq2ad}). These parameters appear implicit in the definition of the variables $X$ and $Y$ and in the $\bar{\gamma}$ parameters (which have small influence in the dynamics of the system). By choosing $k_x=k_y$ the results will be easier to interpret since the fluorescence is directly related to the concentrations of the proteins $x$ and $y$ and, by setting $k_x=k_y$, the fluorescence intensity ratio ($X/Y$ and $Y/X$) and the fluorescence intensity difference ($|X-Y|$) will be directly proportional to these parameter calculated with the real protein concentrations.

An experiment made with the calibrator would consist of cloning a plasmid with the calibrator genetic circuit assembled with the {\it device} promoter (whose parameters one have to choose) among known ones and with a {\it query} promoter whose parameters are unknown. The plasmid should be inserted in cells in solutions of the signaling protein at different concentrations $p_s$. Each promoter is modeled through two parameters, $\bar{\alpha}_{1/2}$ and $k_{1/2}$, $1/2$ stand for {\it device/query} promoter. While at low $p_s$ concentrations both promoters are weak and give a weak fluorescence response, at high $p_s$ concentrations, both promoters are saturated and their strength is maximal. From the fluorescence intensities at these high concentrations of the signaling protein it is possible to establish the relative strength of the two promoters $\bar{\alpha}_2/\bar{\alpha}_1$. In figures \ref{fig:diff} and \ref{fig:rat} we show plots of the fluorescence difference defined as $|X-Y|$ and the fluorescence ratio $X/Y$ for three different values of $\bar{\alpha}_1$ and varying $\bar{\alpha}_2$ at high signaling protein concentrations (the effective strength of both promoters is maximum).

\bfig
\bc
\ig{0.65}{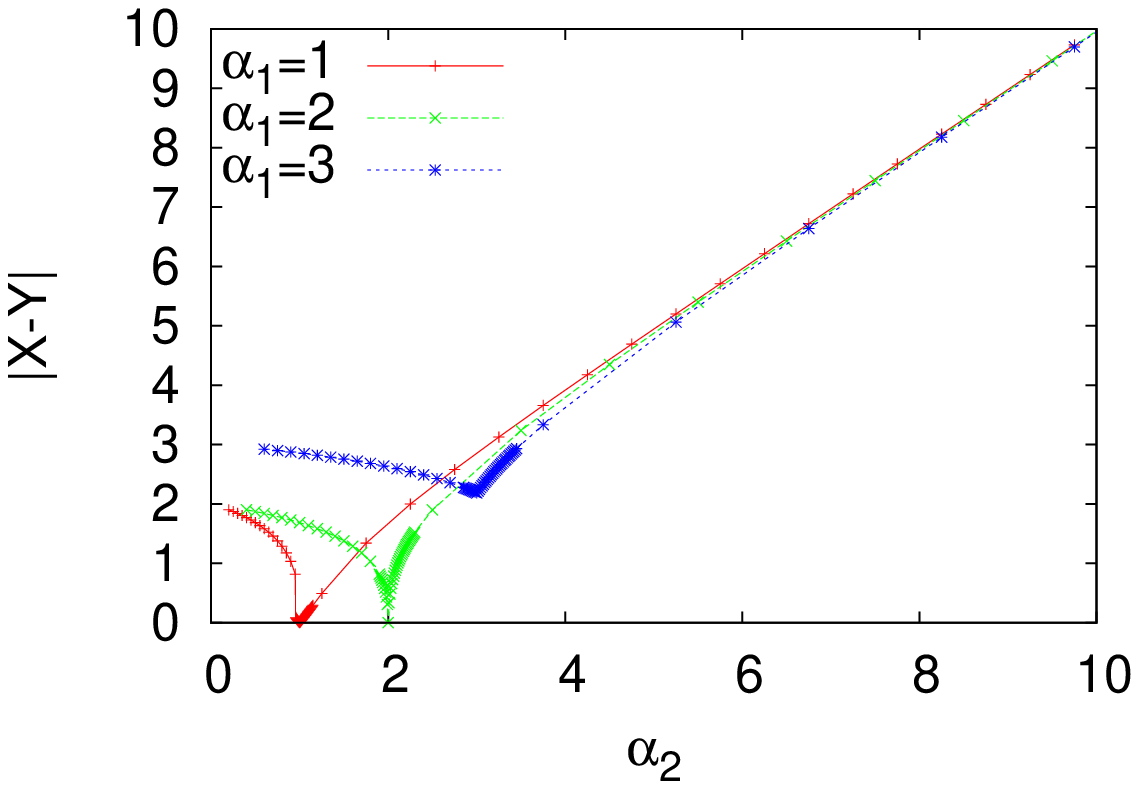}
\caption{(Color online) The fluorescence difference for different values of $\bar{\alpha}_1$ as a function of $\bar{\alpha}_2$. Note that for values of $\bar{\alpha}_2$ sufficiently higher than $\bar{\alpha}_1$ the fluorescence difference increases linearly with the value of $\bar{\alpha}_2$.}\label{fig:diff}
\ec
\efig

\bfig
\bc
\bt{c}
\ig{0.65}{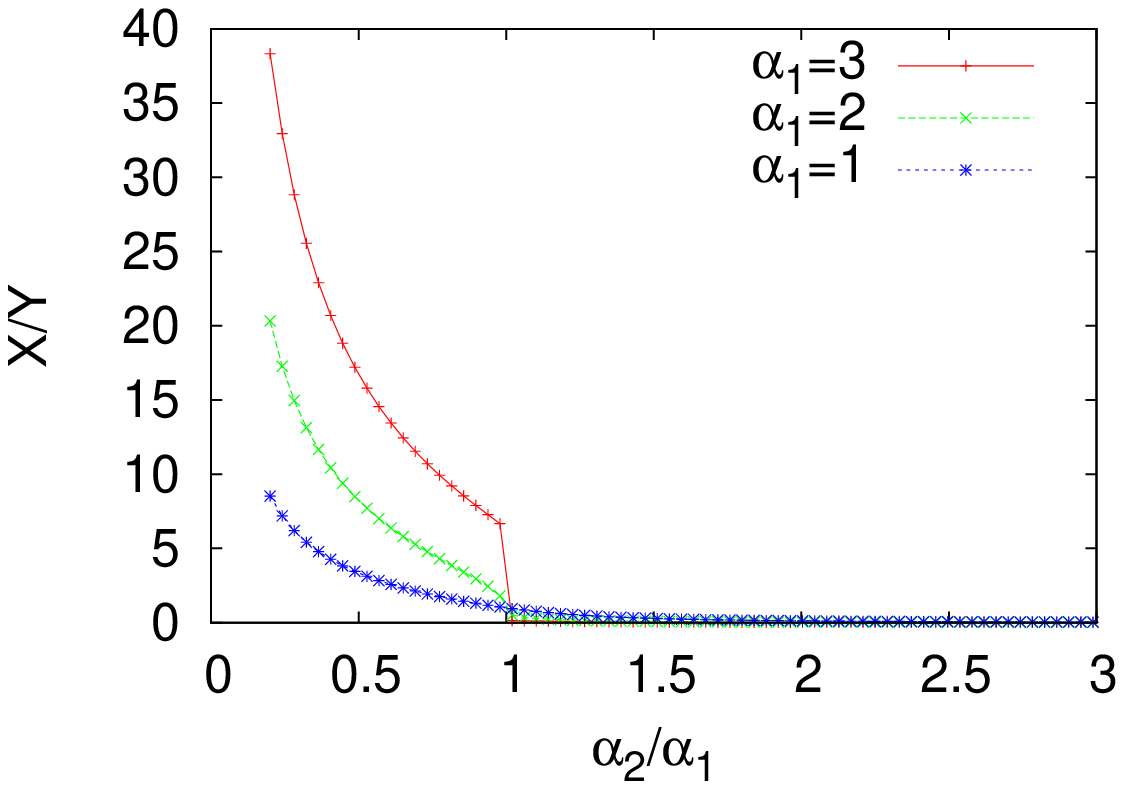} \\ 
\ig{0.65}{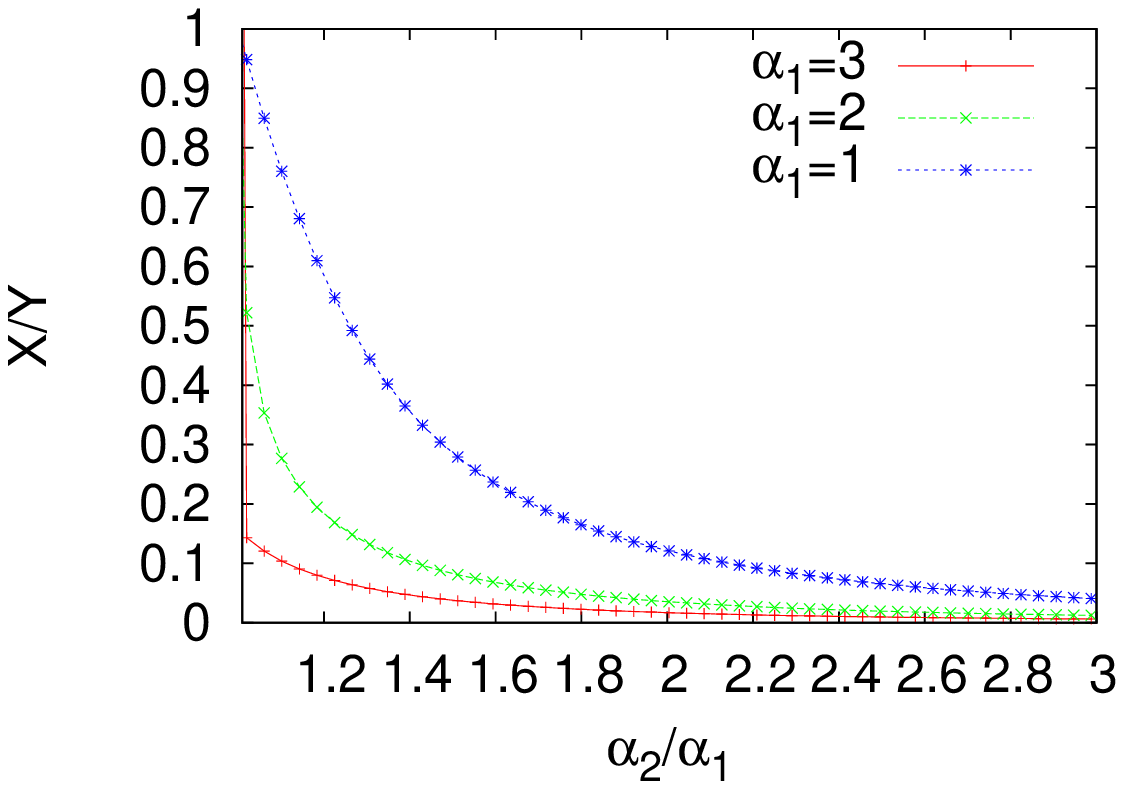}
\et
\caption{(Color online) Upper plot: the fluorescence ratio $X/Y$ as a function of $\bar{\alpha}_2/\bar{\alpha}_1$ for different values of $\bar{\alpha}_1$. One can clearly see that for similar values of $\bar{\alpha}_1$ and $\bar{\alpha}_2$, when the bifurcation occurs, the system goes to a state where the repressor protein of the stronger promoter completely dominates the system. Lower plot: Detail of the region where $\bar{\alpha}_2>\bar{\alpha}_1$.} \label{fig:rat}
\ec
\efig

The first thing to note from figures \ref{fig:diff} and \ref{fig:rat} is that, if the {\it query} promoter is stronger than the {\it device} one, the {\it device} fluorescence ($X$) will be strongly suppressed, and the fluorescence intensity coming from the {\it query} promoter is proportional to its strength (the response of the system is linear). That means, choosing a weak {\it device} promoter, one can establish the relative strength of other promoters by a simple proportionality law given by the linear response plotted in figure \ref{fig:diff}.

At each different $p_s$ concentration, the effective strength of the {\it device} and {\it query} promoters is different, see eq. (\ref{eq7}). The parameter that distinguishes two promoters, with respect to the $p_s$ concentration, is their Michaelis constants, $k_{1,2}$. The parameters $k_{1,2}$ mark the rhythm at which the effective strength of each promoter grows. If a promoter has a small value of $k$, at low $p_s$ concentrations of the signaling protein, the promoter is already acting at full strength, while for high values of $k$ the promoter saturates only at high values of $p_s$. We have already established to choose a small value for the {\it device} promoter $\bar{\alpha}_1$, so we expect the {\it query} promoters to have $\bar{\alpha}_2>\bar{\alpha}_1$. If $k_2<k_1$, the effective strength of the {\it query} promoter is always bigger than the relative strength of the {\it device} one, and in the experiment one observes that the luminosity associated with the {\it query} promoter is stronger for any value of the signaling protein concentration $p_s$. On the other hand, if one chooses a small value for $k_1$, already at low $p_s$ concentrations the strength of the {\it device} promoter saturates, and if $k_1$ is small enough it saturates before the effective strength of the {\it query} promoter reaches a value bigger than $\bar{\alpha}_1$. In this situation one would observe at low concentrations of $p_s$ the luminosity of the {\it device} promoter stronger than the one coming from the {\it query} promoter. Then, at some critical value of $p_{s}=p_{sc}$ both strength are equal and for $p_s>p_{sc}$ the stronger fluorescence is the one from the {\it query} promoter. For $n_1=n_2$, the value of $k_2$ given in units of $k_1$ as a function of $p_{sc}$ (also in units of $k_1$) is given by:

\be
k_2&=&\sqrt[n]{p_{sc}^n\left(\frac{\bar{\alpha}_2}{\bar{\alpha}_1}-1\right)-\frac{\bar{\alpha}_2}{\bar{\alpha}_1}},\label{eq:k2}\\
p_{sc}&=&\sqrt[n]{\left(k_2^n-\frac{\bar{\alpha}_2}{\bar{\alpha}_1}\right)\left(\frac{\bar{\alpha}_1}{\bar{\alpha}_2-\bar{\alpha}_1}\right)}.\label{eq:psc}
\ee

In figure \ref{fig:fluors} we show a few examples of results one might expect for different values of $k_2$.

\bfig[h]
\bc
\bt{c}
\ig{0.55}{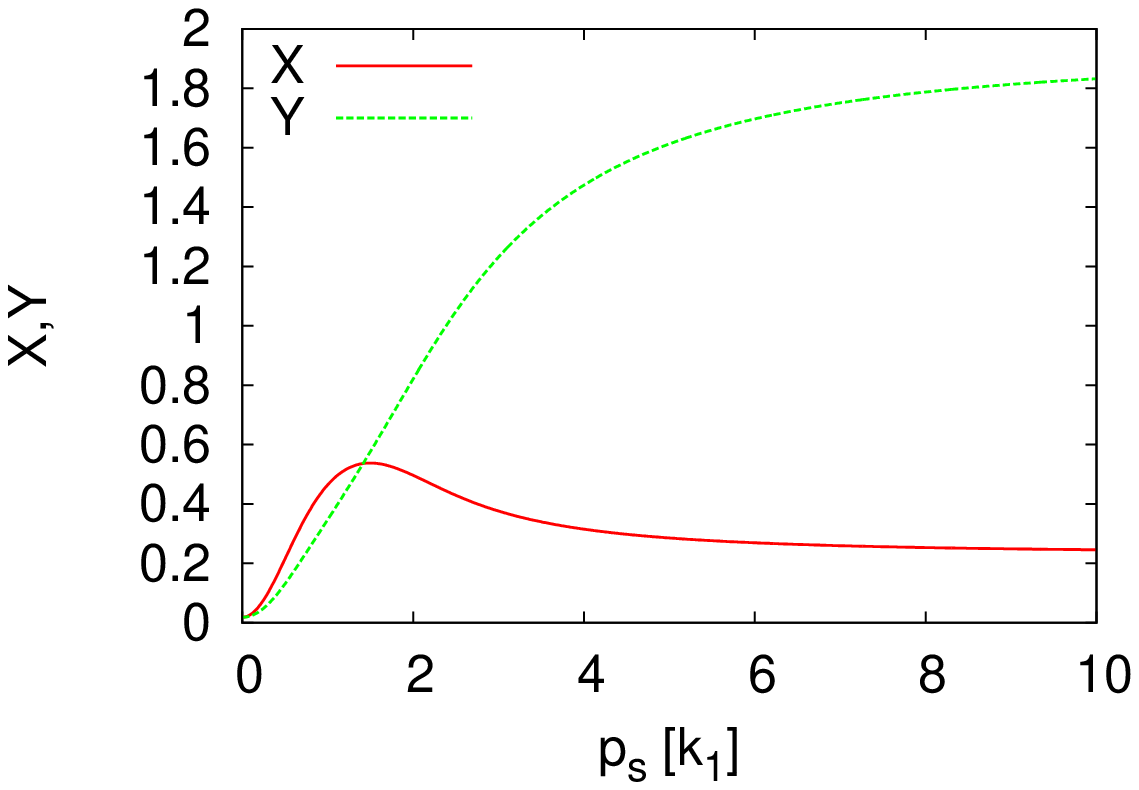} \\\ig{0.55}{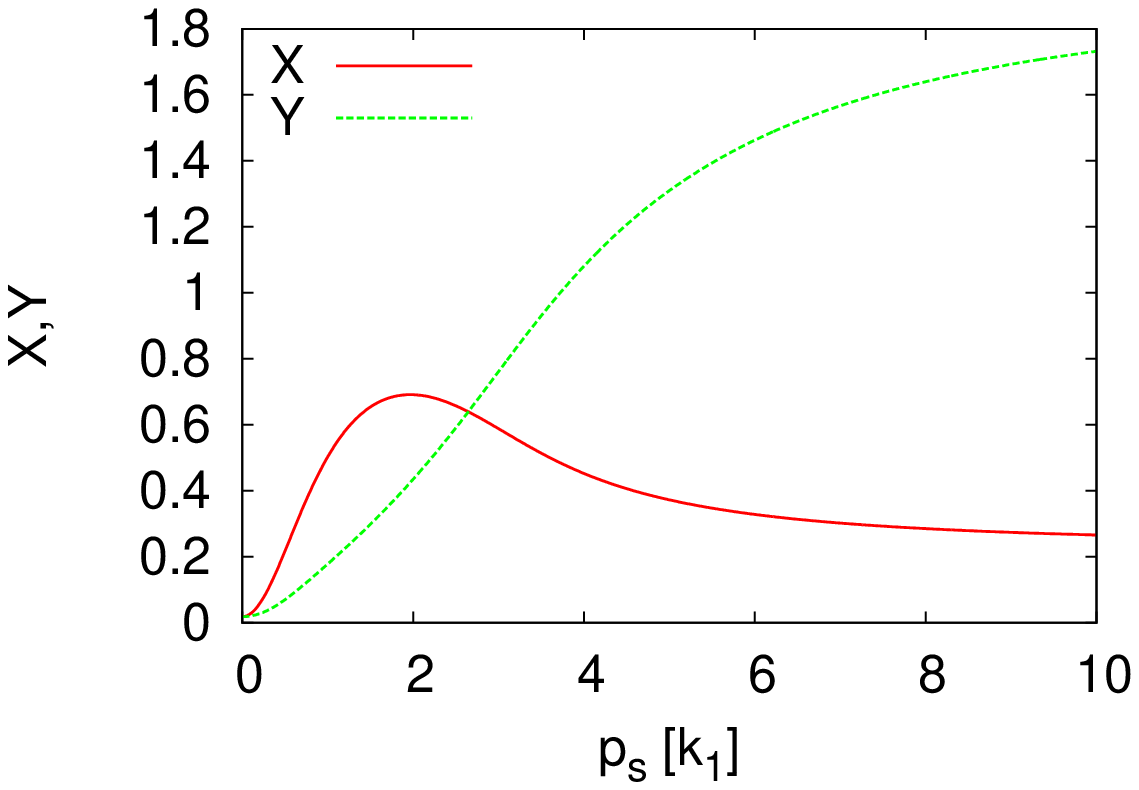} \\\ig{0.55}{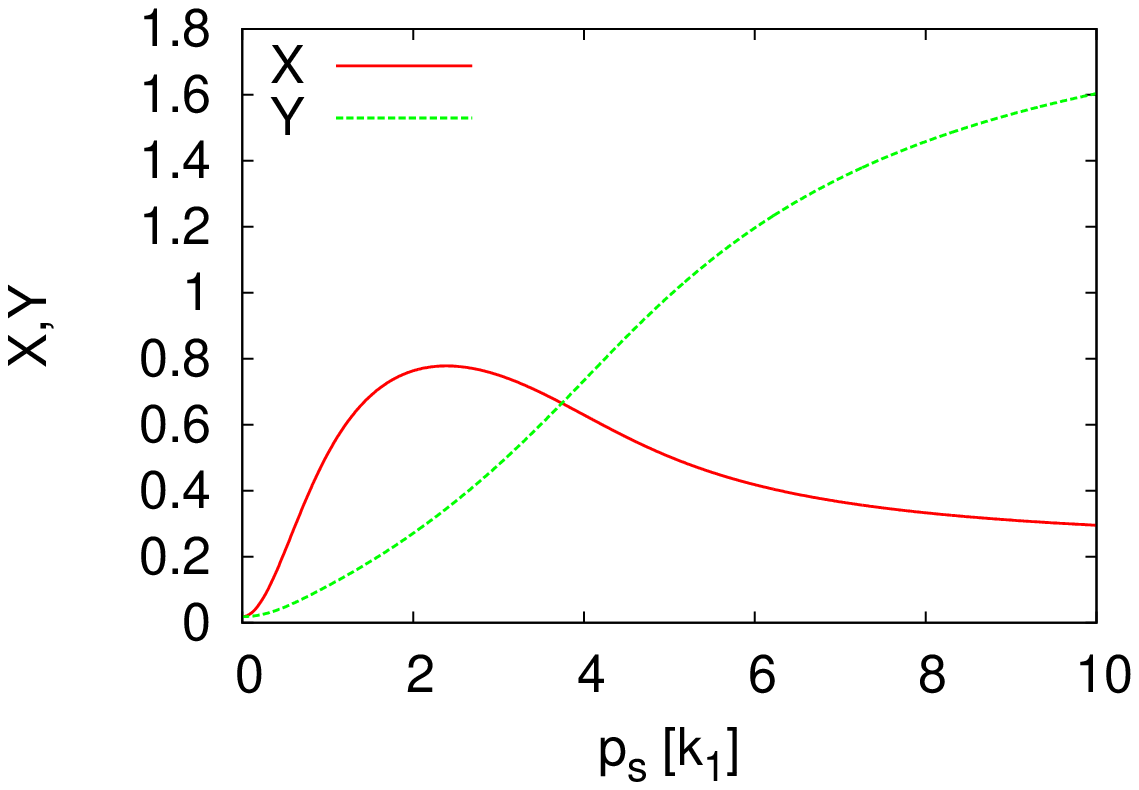} 
\et
\caption{In all plots $\bar{\alpha}_1$=$k_1$=1 and $\bar{\alpha}_2$=2. In the upper plot $k_2$=2, in the center $k_2$=3 and in the bottom plot $k_2$=4. The values for $p_{sc}$ are respectively: $\sqrt{2}$, $\sqrt{7}$ and $\sqrt{14}$.}\label{fig:fluors}
\ec
\efig

\bfig
\bc
\ig{0.65}{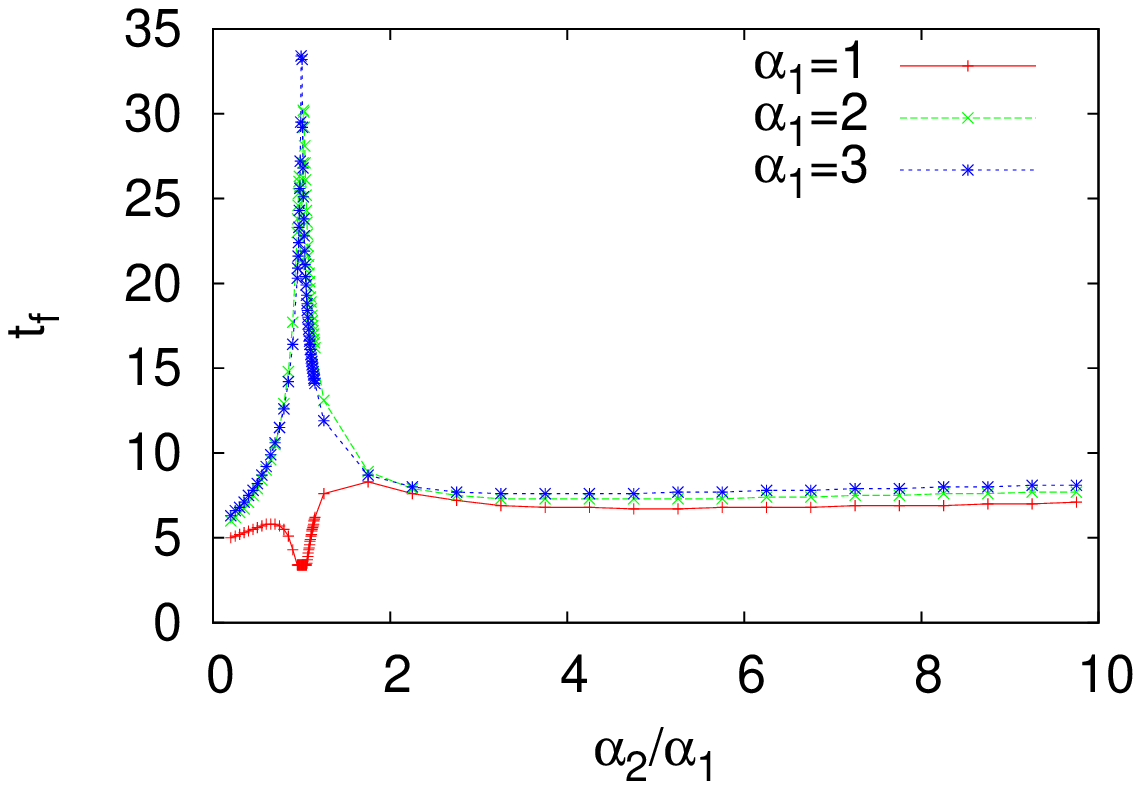}
\caption{Response time of the system for different values of $\bar{\alpha}_1$ and $\bar{\alpha}_2$. For values of the $\bar{\alpha}^\prime$s that the system presents a bifurcation, the response time can be large because the system spend time in its non-equilibrium solution.}\label{fig:temp}
\ec
\efig

So, the construction of the calibrator device, as we present it, would be the following: first one chooses a very weak promoter which has a small Michaelis constant to act as the {\it device} promoter in the calibrator. Second step is to define a standard, to choose a known promoter, clone the calibrator device with it as {\it query} promoter and perform a measurement of the fluorescence intensity of this standard promoter at high $p_s$ concentrations. This fluorescence intensity is the standard one, to which we can compare other promoters. Now performing the experiment with another promoter acting as {\it query} promoter one obtains another value for the luminosity that we can compare with the standard one. The higher or lower this luminosity is with respect to the standard, the stronger or weaker the promoter is compared with the standard, so one can establish the value of $\alpha_2$. Knowing $\alpha_2$ one can perform the same measurement for different $p_s$ concentrations in order to establish the critical value of $p_s$ where the {\it query} fluorescence becomes higher than the {\it device} one. Knowing the value of $p_{sc}$ it is possible to establish the value of $k_2$ by means of eq. \ref{eq:k2} (assuming both promoters have the same $n$).

Now that we have established the ideal parameters for the {\it device} promoter (weak strength and small Michaelis constant) and set $k_x=k_y$ and $\beta_x=\beta_y$ the last important factor is the time response of the system.

In figure \ref{fig:temp} we show plots for the $t_f$, the time the systems needs to reach its steady state\footnote{The system actually goes asymptotically to its steady state without really reaching it. What we have calculated is the time needed so that the sum of the absolute values of the derivatives of $X$ and $Y$ reach a small value (0.01).} for different values of $\bar{\alpha}_1$. One observes that the time response of the system has a peak with the maximum around 30$\beta_x^{-1}$ when the effective strength of both promoters is equal and then it goes to a rather stable value close to 7$\beta_x^{-1}$. For a realistic value of $\beta_x$ like 0.069 min$^{-1}$ the peak value for $t_f$ is 7 hours, while for most of the measurements (the calibrator at different $p_s$ concentrations) this time should be around two hours.


\section{Conclusions}

In the present study we have proposed a biological device that works as a promoter calibrator in which the strength of a collection of {\it query} promoters can be measured against the strength of a {\it device} promoter. Some of the key features of the proposed design are its single cell character, high modularity and handy construction: a unique molecular cloning permits the change of the promoter ready to be calibrated. 
The designed performance of the proposed biological device has been demonstrated by means of an effective mathematical model. The sensitivity analysis of the model shows that there is a sensible relation between the relative promoter strengths and the final steady fluorescence’s measured by the system.

Furthermore, a response time analysis shows that the device can produce a large difference in the repression protein concentrations and in turn in the corresponding fluorescence in approximately two hours.

Finally our promoter calibrator principle may lead to an improvement in the modeling and characterizations of systems in Synthetic Biology, which frequently rely on arbitrarily characterized, or even non-characterized, promoters.


\section*{acknowledgements}
This work has been funded by MICINN TIN2009-12359 project ArtBioCom, the Spanish Ministerio de Educación y Ciencia through the program Juan de la Cierva, the FPI grant program of the Generalitat Valenciana and the Beca de recerca predoctoral from the Universitat Rovira i Virgili.

The authors would also like to thank the Valencia iGEM 2007 team and Enrique O'Connor for useful discussions.



\bibliographystyle{spbasic}

\begin{thebibliography}{99}

\bibitem{alper}
  Alper et al.,
  Tuning genetic control through promoter engineering., PNAS. 102 (36), 12678-12683 (2005).

\bibitem{kumar}
  Kumar A and Snyder M.,
  Genome-Wide Transposon Mutagenesis in Yeast., Current Protocols in Molecular Biology., 13, 13.3 (2001).


\bibitem{elowitz}
  Elowitz MB and Leibler S.,
  A synthetic oscillatory network of transcriptional regulators., Nature, 403, 335-338 (2000).

\bibitem{monod}
  Monod, J. and Jacob,
  General conclusions: teleonomic mechanisms in cellular metabolism, growth and differentiation.,
  Cold spring Harb. Symp. Quant. Biol., 26, 389-401 (1961).

\bibitem{ptashne}
  Ptashne, M.,
  A genetic switch: phage $\lambda$ and Higher Organisms. (1992).

\bibitem{ishiura}
  Ishiura, M et al.,
  Expression of gene cluster kaiABC as a circadian feedback process in cyanobacteria., Science, 281, 1519-1523 (1998)

\bibitem{santos}
  Santos CN and Stephanopoulos G.,
  Combinatorial engineering of microbes for optimizing cellular phenotype., Current Opinion in Chemical Biology., 12, 168-176 (2008)

\bibitem{sayut}
  Sayut DJ,  Niu Y, and Sun L.,
  Construction and Engineering of Positive Feedback Loops., JACS chemical Biology., 1(11), 692-696 (2006)  

\bibitem{weber}
  Weber, W., and Fussenegger, M.,
  Pharmacologic transgene control systems for gene therapy, J. Gene Med., 8, 535–556 (2006).

\bibitem{walz}
  Walz, D., and Caplan, S. R.,
  Chemical oscillations arise solely from kinetic nonlinearity and hence can occur near equilibrium, Biophys. J., 69, 1698–1707 (1995).

\bibitem{gardner}
  Gardner T, Cantor CR and Collins JJ.,
  Construction of genetic toggle switch in Escherichia coli., Nature., 403, 339-342 (2000).

\bibitem{stricker}
   J. Stricker et. al.,
   A fast, robust and tunable synthetic gene oscillator., Nature., 456, 516-519 (2008).

\bibitem{becskei}
  Becskei A and  Serrano L.,
  Engineering stability in gene networks by autoregulation., Nature 405, 590-593 (2000).

\bibitem{becskei2}
  Becskei A, Seraphin B and  Serrano L.,
  Positive feedback in eukaryotic gene networks: cell differentiation by graded to binary response conversion.,
  The EMBO  Journal, 20(10), 2528-2535 (2001).
  
\bibitem{cox}
  Cox, Surette, Elowitz.,
  Programming gene expression with combinatorial promoters., Mol Syst Biol., 3, 145(2007).

\bibitem{miller}
  J. Miller.,
  Experiments in molecular genetics, Cold Spring Harbor Laboratory (1972).

\bibitem{liang}
  Liang et al.,
  Activities of Constitutive Promoters in Escherichia coli., J Mol Biol., 292, 19-37, (1999).

\bibitem{smolke}
  Smolke \& Keasling.,
  Effect of Gene Location, mRNA Secondary Structures, and RNase Sites on Expression of Two Genes in an Engineered Operon.,
  Biotech Bioeng., 80, 762-776 (2002).

\bibitem{khlebnikov}
  Khlebnikov et al.,
  Modulation of gene expression from the arabinose-inducible araBAD promoter., J Ind Microb Biotec., 29, 34-37 (2002).

\bibitem{kelly}
  J. R. Kelly et. al.,
  Measuring the activity of BioBrick promoters using an in vivo standard., Journal of Biological Engineering, 3, 1-13 (2009).

\bibitem{guckenheimer}
  Guckenheimer, J., and Holmes, P.,
  Nonlinear oscillations, dynamical systems, and bifurcations of vector fields., Springer Berlin (1990).


\end{thebibliography}

\end{document}